\documentclass[useAMS,usenatbib]{mn2e}

\usepackage{epsfig}
\usepackage{epstopdf}
\usepackage{lscape} 

\newcommand{\Rin}{R_{\rm in}}
\newcommand{\Mdot}{\dot{M}} 

\title[Nonradiative black hole accretion]{Numerical Parameter Survey of Nonradiative Black Hole Accretion -- Flow Structure and Variability of the Rotation Measure}

\author[Bijia PANG et al.]{
Bijia Pang,$^{1}$\thanks{E-mail:\ bpang@physics.utoronto.ca}
Ue-Li Pen,$^{2}$\thanks{E-mail:\ pen@cita.utoronto.ca}
Christopher D. Matzner,$^{3}$\thanks{E-mail:\ matzner@cita.utoronto.ca}
Stephen R. Green,$^{4}$\thanks{E-mail:\ srgreen@uchicago.edu}
\newauthor 
Matthias Liebend\"orfer,$^{5}$\thanks{E-mail:\ matthias.liebendoerfer@unibas.ch} 
\\
$^1$\,Department of Physics, University of Toronto, M5S 1A7, Canada \\
$^2$\,Canadian Institute for Theoretical Astrophysics, University of
Toronto, M5S 3H8, Canada \\
$^3$\,Department of Astronomy and Astrophysics, University of Toronto, M5S
3H8, Canada \\
$^4$\,Department of Physics, University of Chicago, 5640 South Ellis Avenue, Chicago IL 60637, USA\\
$^5$\,Physics Department, University of Basel, Klingelbergstrasse 82, CH-4056 Basel, Switzerland \\
}

\begin{document}

\date{\today}

\pagerange{\pageref{firstpage}--\pageref{lastpage}} 
\pubyear{2010}

\maketitle

\label{firstpage}

\begin{abstract}

We conduct a survey of numerical simulations to probe the structure
and appearance of non-radiative black hole accretion flows like the
supermassive black hole at the Galactic centre.  We
find a generic set of solutions, and make specific predictions for
currently feasible rotation measure (RM) observations, which
are accessible to current instruments including
the EVLA, GMRT and ALMA.  The slow time variability of the RM is a key
quantitative signature of this accretion flow.  The time variability
of RM can be used to quantitatively measure the nature of the
accretion flow, and to differentiate models.  Sensitive measurements
of RM can be achieved using RM synthesis or using pulsars.

Our energy conserving ideal magneto-hydrodynamical simulations, 
which achieve high dynamical
range by means of a deformed-mesh algorithm, stretch from several 
Bondi radii to about one thousandth of that radius, and continue for
tens of Bondi times.  Magnetized flows which lack outward convection
possess density slopes around -1, almost independent of physical
parameters, and are more consistent with observational constraints
than are strongly convective flows  We observe no tendency for the flows to
become rotationally supported in their centres, or to develop steady
outflow. 

We support these conclusions with formulae which encapsulate our
findings in terms of physical and numerical parameters.  We discuss
the relation of these solutions to other approaches.  The main
potential uncertainties are the validity of ideal MHD and the absence
of a fully relativistic inner boundary condition.  The RM variability
predictions are testable with current and future telescopes.

\end{abstract}

\begin{keywords}
accretion-black hole, MHD simulation
\end{keywords}

\newcommand{\be}{\begin{eqnarray}}
\newcommand{\ee}{\end{eqnarray}}
\newcommand{\beq}{\begin{equation}}
\newcommand{\eeq}{\end{equation}}

\section{Introduction}\label{S:intro} 

The radio source Sgr A* at the Galactic centre (GC) is now accepted to
be a supermassive black hole \citep[$M_{\rm BH}\simeq 4.3 \times
  10^{6} M_\odot$: ][]{2009ApJ...692.1075G}, 
     accreting hot gas
from its environment \citep[$n_e\simeq 130$\,cm$^{-3}, k_B T\simeq
  2$\,keV at 1 arc second:][]{2003ApJ...591..891B}.
Interest in the Sgr A* accretion flow is stimulated by its remarkably
low luminosity; by its similarity to other low-luminosity AGN; by
circumstantial evidence for past episodes of bright X-ray emission
\citep[][but see \citealt{2007ApJ...656..847Y}]{2004A&A...425L..49R}
and nearby star formation \citep{2003ApJ...590L..33L}; and foremost,
by its status as an outstanding physical puzzle.  

Supermassive black holes are enigmatic in many respects; for the GC black hole
(GCBH) the enigma is sharpened by a wealth of observational constraints, which permit detailed, sensitive and spatially resolved studies of
its accretion dynamics.    Within a na\"ive model such as Bondi flow,  matter would flow inward at the dynamical rate from its gravitational sphere of influence, which at $\sim 1''$ is resolved by {\em Chandra}.
Converted to radiation with an efficiency $\eta c^2$, the resulting luminosity would exceed what is actually observed by a factor $\sim 10^5 (\eta/0.1)$.   This wide discrepancy between expectation and observation has stimulated numerous theoretical explanations, including convection \citep{2000ApJ...539..798N,2000ApJ...539..809Q}, outflow \citep{1999MNRAS.303L...1B}, domination by individual stars' winds \citep{2004MNRAS.350..725L}, and conduction \citep{2006ApJ...649..345T,2007ApJ...660.1273J,2008MNRAS.389.1815S,2010arXiv1004.0702S}. 

\subsection{Constraining the accretion flow} 
Because many of its parameters are uncertain, the central density and
accretion rate of the GCBH flow are not strongly constrained by the its
emission spectrum \citep{1999ApJ...520..298Q}; the most stringent
constraints come from observations of the rotation measure
\citep{2000ApJ...545..842Q}, now known to be roughly
$-5.4\times10^5$\,rad\,m$^{-1}$ \citep{2007ApJ...654L..57M}.
Interpreting this as arising within a quasi-spherical flow with
magnetic fields in rough equipartition with gas pressure, and adopting
the typical assumption that magnetic fields do not reverse rapidly, we
derive a gas density $n_H\sim 10^{5.5}$\,cm$^{-3} (R_S/R_{\rm rel})^{1/2}$ at the radius $R_{\rm rel}$ which
dominates the RM integral, namely where electrons become relativistic; see \S \ref{A:RMconstraint} for more detail. 
If this radius is about $10^2$ Schwarzschild radii ($10^2 R_S$), as in the spectral
models of \cite{1999ApJ...520..298Q}, then a comparison between this
density and conditions at the Bondi radius $R_B\simeq 0.053$\,pc
indicates a density power law $\rho\propto r^{-k}$ with 
$k= 1.1-1.3$;the derived value is rather insensitive to the black hole mass, the
degree of equipartition, and the precise radius at which electrons
become relativistic.    (If rapid conduction causes electrons to be nonrelativistic at all radii, the implied slope is falls to 0.8.) 

An independent but weak constraint on $k$ comes from recent
multi-wavelength observations of flares in the emission from Sgr A*.
\cite{2009ApJ...706..348Y} favor an interpretation in which these
flares originate within regions in which electrons have been
transiently heated and accelerated; using equipartition arguments they
estimate a magnetic field strength $B\sim 13-15$\,G at $4-10$
Schwarzschild radii, implying a total pressure $P>20$\,dyn\,cm$^{-2}$
at those radii.  Because $P\propto r^{-(k+1)}$, a comparison to
the conditions at $R_B$ requires $k> 0.6-0.8$.  This constraint could
be violated if the emitting regions were sufficiently over-pressured
relative to the surrounding gas; however the subsonic rate of
expansion inferred by \citet{2009arXiv0907.3786Y} suggests this is not the case.

The density power law $k$ is an important diagnostic, both because it
allows one to estimate the mass accretion rate onto the black hole,
and because $k$ takes definite values within proposed
classes of accretion flows.  \cite{1952MNRAS.112..195B} accretion and
ADAFs \citep[advection-dominated accretion flows,]
[]{1994ApJ...428L..13N}, in which gas undergoes a modified free fall,
imply $k=3/2$ and have long been ruled out \citep{2000ApJ...538L.121A}
by limits on the rotation measure \citep{1999ApJ...521..582B}.  CDAFs
\citep[convection-dominated accretion flows, ][]{2000ApJ...539..798N,2000ApJ...539..809Q}
and related flows like CDBFs \citep[convection-dominated Bondi flows,
][]{2002ApJ...566..137I}, in which convection carries a finite outward
luminosity, all have $k = 1/2$ outside some small radius: otherwise,
convection becomes supersonic \citep{2001astro.ph..4113G}.  

Three classes of flows are known to have intermediate values, $1/2<k<3/2$, as suggested
by the observatons.  One of these is the ADIOS
\citep[advection-dominated inflow-outflow solutions,
][]{1999MNRAS.303L...1B}, in which mass is lost via a wind from all
radii within a rotating ADAF; however these flows appear to require that low angular momentum material has been removed from the axis.  Another is a class of conductive flows, in which heat is carried outward by electrons and stifles accretion at large radii  
\citep{2007ApJ...660.1273J}.  
A third consists of flows which lack any
significant outward convective or conductive luminosity \citep{2001astro.ph..4113G},
but are nevertheless hydrostatic rather than infalling; this behavior
is seen within some numerical simulations in which {\em magnetized}
gas is accreted, such as those of \cite{2003ApJ...592.1042I} and
\cite{2003ApJ...596L.207P}, who termed the flow
``magnetically-frustrated convection''.
 
We are concerned with the last flow class, as it is physically simple,
realizable within simulations, and consistent with observational
constraints.  Whether it is physically relevant depends on the
strength of conduction in the accretion flow, a question we return to
in \S\,\ref{S:discussion}.  Although it is of interest, previous
simulations do not suffice to make any quantitative comparisons
between it and the Sgr A* accretion flow.  \cite{2003ApJ...592.1042I}
have already discussed several shortcomings which afflicted prior
numerical work, such as (1) a lack of energy conservation during
magnetic reconnection and (2) simulation durations too short to
capture steady states or secular trends.  There are a number of other
roadblocks: (3) Dynamical range: $R_B$ is $10^5$ Schwarzschild radii,
but the largest simulations yet done have only a factor of $\sim 10^2$
separating their inner and outer boundaries; (4) Resolution: numerical
solutions are rarely close enough to the continuum limit to allow
turbulent phenomena to be predicted with confidence; (5) Outer
boundary conditions: although matter is presumably fed into the
accretion flow by stellar winds from the nuclear star cluster
\citep{2003ApJ...594..812G}, the flow structure and magnetization of
this gas is not well constrained; (6) Inner boundary conditions: the
hole interacts with the flow in a manner which is not fully
characterized, and which is likely to dominate the energetics; (7)
Mass injection: stars within $R_B$ produce fresh wind material, which
have the potential to affect the final solution
\citep{2004MNRAS.350..725L}; and (8) Plasma physics: close to be black
hole, the flow is only weakly collisional, leading to effects such as
anisotropic pressure and conduction, which may alter the nature of
fluid instabilities and the character of heat transport
\citep{2008MNRAS.389.1815S}.  Potential deviations from ideal MHD
become stronger as one approaches the event horizon, and are discussed
further in section \ref{S:discussion}.

In this paper we describe a numerical parameter survey designed to partially overcome difficulties (1)-(5) in the above list, while making an educated guess regarding (7) and leaving (6) and (8) to future work.   Specifically, we conduct three dimensional, explicitly energy conserving simulations to the point of saturation -- often tens of dynamical times at $R_B$.  We vary the dynamical range and resolution in order to gather information about the astrophysical limits of these parameters, although they lie beyond our numerical reach.  We push numerical outer boundaries far enough from $R_B$ to minimize their effect on the flow, and we vary the conditions exterior to $R_B$ in order to gauge the importance of magnetization and rotation in the exterior fluid.  Our simulations obey ideal MHD, but are viscous and resistive on the grid scale for numerical reasons; we make no attempt to capture non-ideal plasma effects.  We do not account for stellar mass injection within the simulation volume.  Our gravity is purely Newtonian, and at its base we have a region of accretion and reconnection rather than a black hole (although we are currently pursuing relativistic simulations to overcome this limitation).     Our numerical approach is described more thoroughly below. 

By varying the conditions of gas outside $R_B$
and by varying the allocation of grid zones within $R_B$ we are able
to disentangle, to some degree, physical and numerical factors within
our results.  We also compute integrated quantities related to the value and time evolution of RM, and draw conclusions regarding the importance of RM($t$) as a powerful discriminant between physical models. 

We reiterate that our simulations have two simplifications which could
substantially change the behaviour.  1. Our black hole boundary
condition is Newtonian.  Since the deepest potential dominates the
dynamics and energy of the flow, a change in this assumption might
alter the solution.  2. We assume ideal MHD to hold.  As one
approaches the black hole, the Coloumb collision rate is insufficient
to guarantee LTE.  Plasmas can thermalize through other plasma
processes, but if these fail, strong non-ideal effects could dominate
and lead to rapid conduction.  
These effects are both strongest
at small radii, potentially modifying the
extrapolation to the actual physical parameters.  We address these
issues in more detail in section \ref{S:discussion}.

\section{Simulation}\label{S:Simulation} 

\subsection{Physical setup and dimensionless physical parameters}\label{SS:setup+params}

We wish our simulations to be reasonably realistic with regard to the
material which accretes onto the black hole, but also easily described
by a few physical and numerical parameters.  We therefore do not treat
the propagation and shocking of individual stellar winds or turbulent
motions, but take the external medium to be initially of constant
density $\rho_0$ and adiabatic sound speed $c_{s0}$, and imbued with a
characteristic magnetic field $B_0$ and characteristic rotational
angular momentum $j_0$ (but no other initial velocity).  A Keplerian
gravity field $-G M/r^2$ accelerates material toward a central ``black
hole'' of mass $M$ surrounded by a central accretion zone.  The Bondi accretion radius is therefore
\begin{equation} \label{R_B}
R_B=\frac{G M}{c_{s0}^2}. 
\end{equation} 
We adopt the Bondi time $t_B = R_B/c_{s0}$ as our basic time unit; this is 100 years for the adopted conditions at Sgr A*.  All of
the initial flow quantities will evolve as a result of this during the
course of the simulation, and we run for many Bondi times in order to
allow the accretion flow to settle into a final state quite different
from our initial conditions.  From the above dimensional quantities we 
define several dimensionless physical parameters.

The adiabatic index is $\gamma=5/3$; the initial plasma-$\beta$
parameter, or ratio of gas to magnetic pressure, is 
\begin{equation}\label{beta}
\beta_0 = {8\pi \gamma \rho_0 c_{s0}^2\over B_0^2};
\end{equation} 
we consider models with $\beta_0 = (1,10,100,1000,\infty)$  to
capture a wide range of plausible magnetizations.  In our main
sequence of simulations we adopt a uniform magnetic field ${\mathbf
  B}_0$.\footnote{ We also investigated scenarios with Gaussian random
  field components, in which the dominant wavelengths were some
  multiple of $R_B$; however we abandoned these, as such fields decay
  on a Alfv\'en crossing time, confounding our attempts to quantify
  the accretion flow, and we did not wish to add a turbulent driver to
  maintain steady state.}

The initial velocity field is $\mathbf{v}_0 = ({\mathbf j}_0 \times
\hat{\mathbf r})/r$, where $\mathbf r$ is the separation from the
black hole.  The specific vector angular momentum is thus ${\mathbf
  j}_0$ at the rotational equator, with solid-body rotation on
spherical shells away from the equator.  A dimensionless rotation
parameter is therefore
\begin{equation}
{R_K\over R_B} = \left(j_0 c_s \over G M\right)^2; 
\end{equation} 
here $R_K = j_0^2/(GM)$ is the Keplerian circularization radius of the
equatorial inflow.  (Our flows never do circularize at $R_K$, both because
angular momentum transport alters the distribution of $j$, and because
gas pressure can never be neglected.)

We impose mass accretion and magnetic field reconnection within a zone
of characteristic radius $\Rin$, described below, which introduces the
dynamic range parameter $ R_B/\Rin$.  Because it sets the separation
between small and large scales and the maximum depth of the potential
well, this ratio has a strong influence on flow properties.  One of
our goals is to test how well the flow quantities at high dynamic
range can be predicted from simulations done at lower dynamic range,
as the dynamic range appropriate to Sgr A* is beyond what we can
simulate.

\begin{figure*}
\centering
\includegraphics[scale=0.9]{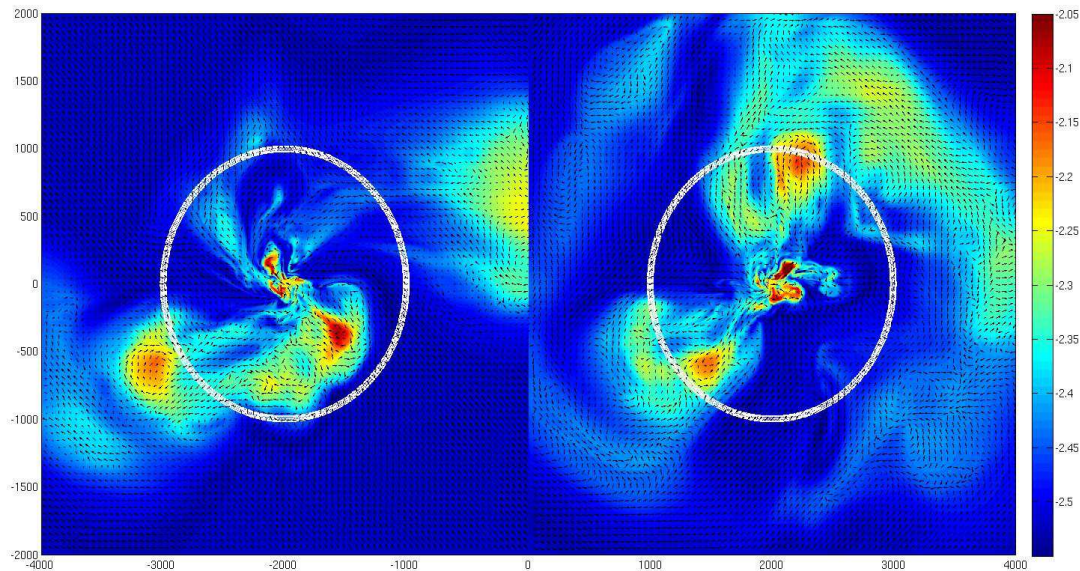}
\caption{2D slice of the simulation for $600^3$ box at 15 Bondi times.
Colour represents the entropy,
and arrows represent the magnetic field vector.
The right panel is the equatorial plane (yz), while the
left panel a perpendicular slice (xy).
White circles represent the Bondi radius ($r_B=1000$).
The fluid is slowly moving, in a state of magnetically frustrated convection.
A movie of this flow is available in the supporting
information section of the electronic edition.
}
\label{figure:2D_rK}
\end{figure*}

\subsection{Grid setup and numerical parameters}

We employ a fixed, variable-spacing Cartesian mesh in which the grid
spacing increases with distance away from the black hole.  To simplify
our boundary conditions, we hold the spacing fixed within the inner
accretion zone and near the outer boundary.  The total box size is
$4000^3$ in units of the minimum grid spacing; however this is
achieved within a numerical grid of only $300^3$ 
to $600^3$
zones.  Our grid geometry allows for a large number of long-duration runs to be
performed at respectable values of the dynamic range, while avoiding
coordinate singularities and resolution boundaries.  These advantages
come at the cost of introducing an anisotropy into the grid resolution; however we have tested the code for
conservation of angular momentum and preservation of magnetosonic
waves, and found it to be comparable in accuracy to fixed-grid codes
with the same resolution.  Our grid expansion factor $s =
\delta dx_i/dx_i$ takes one value for $x_i<R_B$ and another, larger
value for $x_i > R_B$; this allows us to devote most of our
computational effort to the accretion region of interest, while also
pushing the (periodic) outer boundary conditions far away from this
region.  The inner expansion factor $s_{\rm in}$ is therefore an
important numerical parameter, related to both the grid's resolution
and its anisotropy where we care most about the flow. 

Within our inner accretion region, magnetic fields are reconnected
(relaxed to the vacuum solution consistent with the external field, see
appendix \ref{A:InnerBCs})
and mass and heat are adjusted (invariably, removed) so that the sound
speed and Alfv\'en velocity both match the Keplerian velocity at
$R_B$.  The accretion zone is a cube, whose width we hold fixed at 15
in units of the local (uniform) grid separation, so we define $\Rin =
7.5\, dx_{\rm min}$ (but note, the volume of this region is equivalent
to a sphere of radius $9.3 dx_{\rm min}$.)  We consider it too costly
to vary the numerical parameter $\Rin/dx_{\rm min}$.  

Our grid geometry imposes a local dimensionless resolution parameter
\begin{equation}\label{Re-param}
\Re \equiv {r\over \max_i(dx_i)}
\end{equation}
(the maximum being over coordinate directions), which depends both on
radius and on angle within the simulation volume.  At the inner
boundary $\Re \simeq 7.5-9.3$; $\Re$ increases to nearly $s_{\rm
  in}^{-1} \simeq 10^2$ at $R_B$, then decreases toward $s_{\rm
  out}^{-1}$ in the exterior region.  In \S \ref{S:results} we report
the effective resolution at the Bondi radius, $\Re_B = \Re(R_B)$,
along with our results.

\subsection{Computational implementation}\label{implementation} 

Our simulations were performed on the Canadian Institute for
Theoretical Astrophysics Sunnyvale cluster: 200 Dell PE1950 compute
nodes; each node contains 2 quad core Intel(R) Xeon(R) E5310 @ 1.60GHz
processors, 4GB of RAM, and 2 gigE network interfaces.  The code
\citep{2003ApJS..149..447P} is a second-order accurate (in space and
time) high-resolution total variation diminishing (TVD) MHD parallelized
code.  Kinetic, thermal, and magnetic energy are conserved and
divergent of magnetic field was kept to zero by flux constrained
transport.  There is no explicit magnetic and viscous dissipation in
the code except on the grid scale.  
We used an MPI version \citep{2009arXiv0910.2854K},
and 216 CPU cores were used to compute a
$300^{3}$ box, using OpenMP with 8 cores per node.
The $600^3$ simulation was performed on the SciNet cluster using
1000 cores over 125 nodes \footnote{http://www.scinet.utoronto.ca/}.

A parallel effort to implement the code on economic graphics processing
units (GPUs) is in progress, which will allow larger and longer
simulations in the future \citep{2010arXiv1004.1680P}. 

\section{Simulations and Results}\label{S:results}

\begin{figure}
\centering
\includegraphics[scale=0.4]{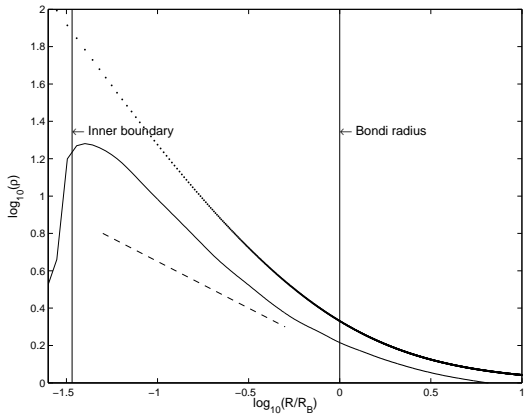}
\caption{Density versus radius.
The dotted line represents the density profile for the Bondi solution,
which is the steepest plausible slope at $k=1.5$.
The dashed line represents the density scaling for CDAF solution,
which is the shallowest proposed slope with $k=0.5$.
The solid line is the density profile from one of our simulations,
which is intermediate to the two.
}
\label{figure:radial}
\end{figure}

Our suite of simulations is described in Table \ref{table:Mdot}, along
with some selected results. We independently varied the magnetization,
rotation, and dynamic range of the flow, as well as the effective
resolution at $R_B$.  In order to suppress the lingering effects of
our initial conditions, we ran each simulation for long enough that a
total mass equivalent to all the matter initially within $R_B$ was
eventually accreted, before assessing the flow structure.  Because
most of our runs exhibited a significant suppression of the mass
accretion rate $\Mdot$ relative to the Bondi value, this constraint
required us to simulate for many $t_B$ (typically 20 $t_B$).  This
requirement put strenuous constraints on our simulations (each of
which required $\sim 3$ weeks to complete), and will be a serious
limitation on any future simulations performed at higher dynamic
range.

\begin{table*}
\centering
\begin{minipage}{\textwidth}
\centering
\begin{tabular}{|c|c|c|c|c|c|c|c|c|c|}  
\hline
Run & $R_B\over dx_{\rm min}$ & $R_B\over \Rin$ & 1+$s_{\rm in}$ &  $\Re_B$  & $\beta_0$ & $R_K\over R_B$ & $t_{\rm sim}\over t_B$ & $\dot{M}\over\dot{M}_{Bondi}$ & $k_{\rm eff}$\footnote{Values are taken from Equation \ref{eqn:keff}} \\
\hline
1  & 500 & 67 & 1.023 & 40.15 & $\infty$ & 0 & 8 & 1.02 & 1.5047 \\
2  & 250 & 33 & 1.013 & 59.29 & $\infty$ & 0 & 3 & 1.10 & 1.5273 \\
3  & 125 & 17 & 1.013 & 48.11 & 100 & 0 & 6-20 & 0.49 & 1.2482 \\
4  & 250 & 33 & 1.013 & 59.29 & 100 & 0 & 6-20 & 0.31 & 1.1650 \\
5  & 500 & 67 & 1.023 & 40.15 & 100 & 0 & 6-20 & 0.22 & 1.1399 \\
6  &1000 & 133 & 1.0315 & 30.82 & 100 & 0 & 6-10 & 0.16 & 1.1253 \\
7  & 250 & 33 & 1.013 & 59.29 & 1 & 0 & 6-20 & 0.15 & 0.9574 \\
8  & 250 & 33 & 1.013 & 59.29 & 10 & 0 & 6-20 & 0.26 & 1.1147 \\
9  & 250 & 33 & 1.013 & 59.29 & 1000 & 0 & 6-20 & 0.40 & 1.2379 \\
10 & 250 & 33  & 1.013 & 59.29 & 100 & 0.1 & 6-20 & 0.289 & 1.1450 \\
11 & 250 & 33  & 1.013 & 59.29 & 100 & 0.5 & 6-20 & 0.286 & 1.1420 \\
12 & 250 & 33  & 1.013 & 59.29 & 100 & 1.0 & 6-20 & 0.31 & 1.1650 \\
13\footnote{case of $75^3$ grid resolution} & 62.5 & 33 & 1.06 & 14.24 & 100 & 0 & 6-20 & 0.30 & 1.1557 \\
14\footnote{case of $150^3$ grid resolution} & 125 & 33 & 1.037 & 28.94  & 100 & 0 & 6-20 & 0.33 & 1.1829 \\
15 & 250 & 33 & 1.013 & 59.29 & $\infty$ & 0.1 & 6-20 & 0.615 & 1.3610 \\
16 & 250 & 33 & 1.013 & 59.29 & $\infty$ & 0.5 & 6-20 & 0.621 & 1.3637 \\
17 & 250 & 33 & 1.013 & 59.29 & $\infty$ & 1.0 & 6-20 & 0.759 & 1.4211 \\
18 & 250 & 33 & 1.013 & 59.29 & 1000 & 0.1 & 6-20 & 0.400 & 1.2379 \\
19\footnote{19-23, B field is along [0 0 1] axis} & 250 & 33 & 1.013 & 59.29 & 1000 & 0.1 & 6-20 & 0.469 & 1.2835 \\
20 & 250 & 33 & 1.013 & 59.29 & 100 & 0.1 & 6-20 & 0.300 & 1.1557 \\
21 & 250 & 33 & 1.013 & 59.29 & 10 & 0.1 & 6-20 & 0.233 & 1.0834 \\
22 & 250 & 33 & 1.013 & 59.29 & 1 & 0.1 & 6-20 & 0.188 & 1.0220 \\
23 & 250 & 33 & 1.013 & 59.29 & 100 & 0 & 6-20 & 0.340 & 1.1915 \\
24\footnote{24-25, B field is along [1 2 0] axis} & 500 & 67 & 1.0315 & 31.65 & 100 & 0.1 & 6-20 & 0.18 & 1.2434 \\ 
25\footnote{case of $600^3$ grid resolution} & 1000 & 58.9 & 1.015 & 64 & 100 & 0.1 & 6-20 & 0.19 & 1.0925 \\ 
\hline
\end{tabular}
\end{minipage}
\caption{Simulations described in this paper.  Columns: Run number; Maximum resolution relative to the Bondi radius; Radial dynamic range within $R_B$; grid expansion factor within $R_B$; effective resolution at $R_B$; magnetization parameter; rotation parameter; range of simulation times over which flow properties were measured; mean mass accretion rate over this period; and typical density power law slope ($\rho\propto r^{-k}$) over this period.
}
\label{table:Mdot}
\end{table*}

\subsection{Character of saturated accretion flows}\label{SS:Results:Character}

\begin{figure}
\centering
\includegraphics[scale=0.4]{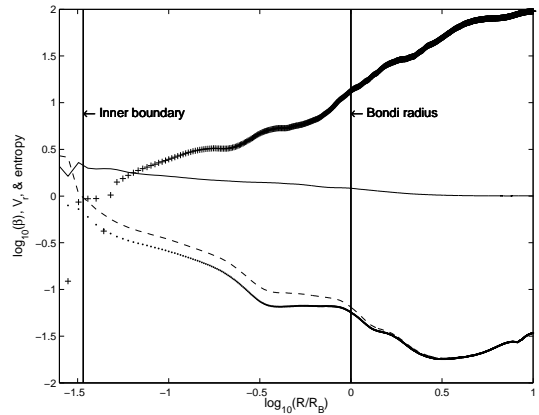}
\caption{$\log({\beta})$, entropy and radial velocity versus radius.
The dashed line $v_r/c_s$ 
represents the radial velocity in units of mach number.
The dots $v_r/c_{ms}$ represent the radial velocity in units of magnetosonic 
mach number. The solid line is the entropy, and we see the entropy
inversion which leads to the slow, magnetically frustrated convection.
Inside the inner boundary, the sound speed is lowered, leading to the
lower entropy.  The + symbols are the magnetic field strength, $\beta$.
}
\label{figure:beta_Vr_radius}
\end{figure}

Figure \ref{figure:2D_rK} shows the 2D slices for the simulation of our
highest resolution $600^3$ box at 15 Bondi times 
(case 25) \footnote{Movies are also available in various
formats at http://www.cita.utoronto.ca/~pen/MFAF/blackhole{\_}movie/index.html}.
The remaining Figures are all based on case 10, which is most representative
of the whole set of simulations.
Figures \ref{figure:radial} and \ref{figure:beta_Vr_radius} display the 
spherically-averaged properties,
figure \ref{figure:radial} shows the spherically-averaged
density of the run;
figure \ref{figure:beta_Vr_radius} shows the spherically-averaged radial 
velocity, $\beta$ and entropy (normalized to the Bondi entropy).  
The entropy inversion is clearly visible,
which leads to the slow, magnetically frustrated convection.

We draw several general conclusions from the runs listed in Table \ref{table:Mdot}: 
\begin{itemize}
\item[-] In the presence of magnetic fields, the flow develops a super-adiabatic temperature gradient and flattens to $k\sim 1$.   Gas pressure remains the dominant source of support at all radii, although magnetic forces are always significant at the inner radius. 
\item[-] Mass accretion diminishes with increasing dynamic range, taking values $\Mdot\simeq (2-4)\Mdot_B (\Rin/R_B)^{3/2-k}$. 
\item[-] Even significant rotation at the Bondi radius has only a minor impact on the mass accretion rate, as the flows  do not develop rotationally supported inner regions. 
\item[-] Our results depend only weakly on the effective resolution $\Re_B$. 
\item[-]  In the absence of magnetic fields and rotation, a Bondi flow develops.   (\citealt{2003ApJ...596L.207P} further demonstrated a reversion to Bondi inflow if magnetic fields are suddenly eliminated; we have not repeated this experiment.)
\end{itemize}

\subsubsection{Lack of rotational support} \label{SSS:norotation}

The non-rotating character of the flow casts some doubt on models
which depend on equatorial inflow and axial outflow.  Our
nonrelativistic simulations cannot rule out an axial outflow from a
spinning black hole, but they certainly show no tendency to develop
rotational support in their inner regions, even after many tens of
dynamical times.  In a rotating run, angular momentum is important at
first, in preventing the accretion of matter from the equator.  Axial,
low-$j$ material does accrete, but some of it shocks and drives an
{\em outflow} along the equator (as reported by
\citealt{2003ApJ...596L.207P} and \citealt{2003ApJ...592..767P}).  After a few $t_B$
this quadrupolar flow disappears, leaving behind the nearly
hydrostatic, slowly rotating envelope which will persists for 
our entire simulation time, i.e. tens of
$t_B$.  We attribute the persistence of this rotational profile to
magnetic braking, as the Alfv\'en crossing time of the envelope is
always shorter than its accretion time.  Magnetic fields thus play a
role here which is rather different than in simulations which start
from a rotating torus, where the magneto-rotational instability is the
controlling phenomenon; the critical distinction is the presence of
low-angular-momentum gas.

Unlike compact object disks, which accrete high-angular-momentum material and are guaranteed to cool in a fraction of their viscous time, the GCBH feeds upon low low-angular-momentum matter, and its accretion envelope cannot cool.   For both of these reasons it is not surprising to discover a thick, slowly rotating accretion envelope rather than a thin accretion disk. 
We stress that global simulations, which resolve the Bondi radius
and beyond and continue for many dynamical times, are required to capture the
physical processes which determine the nature of the flow.

\subsubsection{Dependence on parameters; Richardson extrapolation} \label{SSS:Paramdep+extrap}

We now investigate whether our results for the accretion rate can be distilled into a single, approximate expression.  It is clear from the results in Table \ref{table:Mdot} that rotation affects the accretion rate in a non-monotonic fashion.  However as we have just noted that rotation plays a minor role in our final results, we are justified in fitting only the non-rotating runs.    Rather than $\dot{M}/\dot{M}_{\rm Bondi}$ we fit an effective density slope $k_{\rm eff}$ defined by 
\begin{equation}
\frac{\dot{M}}{\dot{M}_{Bondi}}=\left({R_{\rm in} \over R_{B}}\right)^{3/2 - k_{\rm eff}}.
\label{eqn:keff}
\end{equation}
There are four major variables:
the magnitute of the ambient magnetic field (${\beta_{0}}$),
the radial dynamical range (${R_{B}}/{R_{\rm in}}$),
the resolution of the Bondi scale  ($\Re_B$).  Our fit is 
\begin{equation}
{k_{\rm eff}}=1.50-0.56\beta_{0}^{-0.098}+6.51\left({R_{B}\over R_{\rm in}}\right)^{-1.4}-0.11\Re_B^{-0.48}; 
\label{eqn:kfit}
\end{equation}
all seven numerical coefficients and exponents were optimized against the 25 runs in Table \ref{table:Mdot} . 
The form of equation (\ref{eqn:kfit}) is significantly better than others we tested, including those involving
$\log({R_{B}}/{R_{\rm in}})$ and $\log(\Re_B)$. 
It predicts the entries in Table \ref{table:Mdot} to within a root-mean-square error of only 0.017. 

Somewhat unexpectedly, this nonlinear fit to our simulation output recovers the Bondi solution in the continuum, unmagnetized limit ($k_{\rm eff}\rightarrow 3/2$ as $\beta_0\rightarrow\infty$, $R_B/R_{\rm in}\rightarrow \infty$, and $\Re_B\rightarrow\infty$).  Moreover the form of the expression allows us to extrapolate, in the manner of Richardson extrapolation, to conditions we expect are relevant to Sgr A*: $\Re_B \sim \infty$, $R_B/R_{\rm in} \sim 10^5$, and $\beta_0 \sim 1-5$: then, $k_{\rm eff} \sim 0.94-1.0 $.   

It is encouraging that this result lies in the vicinity of
observational constraints, lending additional credence to the notion
that Sgr A* is surrounded by a ``magnetically-frustrated'' accretion
flow.   We must recall, however, that this is only an extrapolation
based on simulations which lack potentially important physics such as
a relativistic inner boundary and a non-ideal plasma.  The absence of
an imposed outward convective luminosity is likely to be the essential
element which allows for a lower value of $k$. 

\section{Rotation measure} \label{S:RM}

The magnitude of RM constrains the density of the inner accretion flow, thereby also constraining the mass accretion rate and power law index $k$.   Future observations should provide time series of RM$(t)$, a rich data set which encodes important additional information about the nature of the flow.    Our goal in this section will be to characterize RM variability within our own simulations sufficiently well to distinguish them from other proposed flow classes. 

We pause first to consider why RM should vary at all.  The rotation of polarization is determined by an integral (eq. \ref{RM}, \citealt{2008ApJ...688..695S}) which is proportional to $\int n_e {\mathbf B}\cdot {\mathbf dl}$ integrated over the zone of nonrelativistic electrons.   The integral is typically dominated by conditions at $R_{\rm rel}$, the radius where $kT_e = m_e c^2$.   Even if $n_e$ is reasonably constant, ${\mathbf B}$ likely will change in magnitude and direction as the flow evolves.  Given that the dynamical time at $R_{\rm rel}$ is under a day, any strongly convective flow should exhibit significant day-to-day fluctuations in RM; measurements by \cite{2007ApJ...654L..57M} appear to rule this out.   Rotational support also implies rapid RM fluctuations unless $\mathbf B$ is axisymmetric.  In the highly subsonic flow of magnetically-frustrated convection, however, RM may vary on much longer time scales. 

Two  proposals have been advanced in which RM($t$) would be roughly constant.  Within their simulations of thick accretion disks,  \cite{2007ApJ...671.1696S} show that trapping of poloidal flux lines leads to a rather steady value of RM for observers whose lines of sight are out of the disk plane.  \cite{2008MNRAS.389.1815S} point to the constancy of RM in the steady, radial magnetic configuration which develops due to the saturation of the magneto-thermal instability (in the presence of  anisotropic electron conduction).  We suspect that noise at the dynamical frequency 
is to be expected 
in both these scenarios, which need not exist in a magnetically frustrated flow.  We also note that both scenarios lead to systematically low values of RM for a given accretion rate, and therefore imply somewhat higher densities than we inferred from a spherical model; this may be observationally testable.  

\begin{figure}
\centering
\includegraphics[scale=0.4]{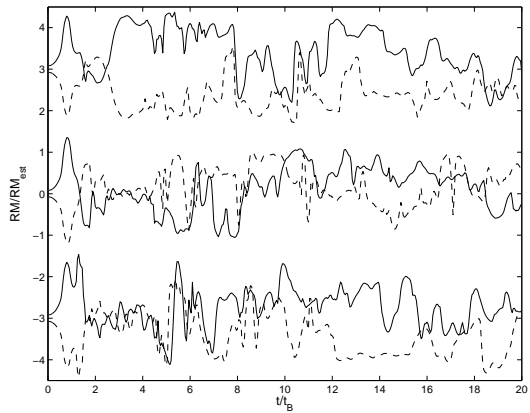}
\caption{Rotation measure vs time (in units of $t_B$).
We chose $R_{\rm rel} = 17$, corresponding to
$R_{\rm rel}/R_{\rm B}$=0.068.
Six lines represent three axes:
upper set is X (centered at +3), center is Y (centered at 0) and lower
is Z (centered at -3),
with positive and negative directions drawn as solid and dashed
lines, respectively.
}
\label{figure:RMvsTime}
\end{figure}

\begin{figure}
\centering
\includegraphics[scale=0.4]{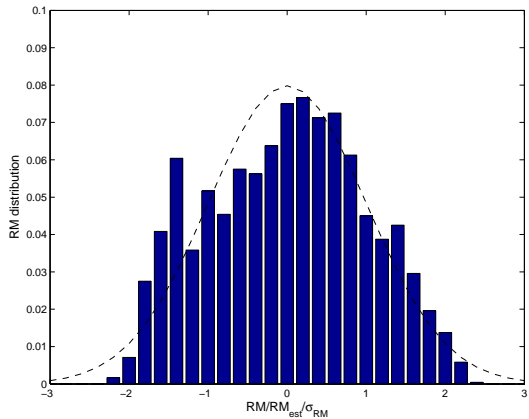}
\caption{PDF of RM in Figure~\protect\ref{figure:RMvsTime}.
The dashed line represents a Gaussian distribution.
The horizontal axis has been normalized by the standard deviation
in figure \protect\ref{figure:RMvsTime}, ${\sigma_{\rm RM}}=0.63$.
}
\label{figure:RM_PDF}
\end{figure}

Our calculation of RM($t$) is based on case $10$ in Table
\ref{table:Mdot}. In Figure \ref{figure:RMvsTime} we plot RM($t$)
against an analytical estimate of its magnitude.  For this purpose we
estimate RM as,
\begin{equation}
{\rm RM} \equiv \frac{e^3} {2\pi {m_e}^2 {c^4}} \int_{R_{\rm rel}}^{R_B} {n_e} B dr, 
\end{equation}
integrated along radial rays (two per
coordinate axis) through the simulation volume.  We neglect the
difference between this expression and one which accounts for the
relativistic nature of electrons within $R_{\rm rel}$.  We therefore
normalize RM to the estimate RM$_{\rm est}$ as,
\begin{equation} \label{RM_est} 
{\rm RM}_{\rm est}= {e^3 \over 2 c^4 m_e^2} \left[ GM R_{\rm rel} \mu_e n_e(R_{\rm rel})^3 \over 11\pi \right]^{1/2} 
\end{equation}
given by equation (\ref{RMform})
with $F(k,k_T)\rightarrow 1$, $\left<\cos(\theta)\right>\rightarrow1/2$,
$\beta\rightarrow10$, and $k\rightarrow1$.  Because we do not calculate electron
temperature within our simulations, we have the freedom to vary
$R_{\rm rel}$ and to probe the dependence of coherence time on this
parameter.  In practice we chose $R_{\rm rel} = (17, 26, 34, 43)
\delta x_{\rm min}$ in order to separate this radius from the
accretion zone ($7.5 \delta x_{\rm min}$) and Bondi radius ($250
\delta x_{\rm min}$, in this case).  Figure \ref{figure:RMvsTime}
illustrates RM$(t)$ along each coordinate axis with the case for
$R_{\rm rel}=17\delta x_{\rm min}$).
As this figure shows, RM changes slowly and its amplitude agrees with our estimate $RM_{\rm est}$.  In our simulations, we can measure the
full PDF, shown in figure~\ref{figure:RM_PDF}.

\begin{figure*}
\centering
\includegraphics[scale=1.0]{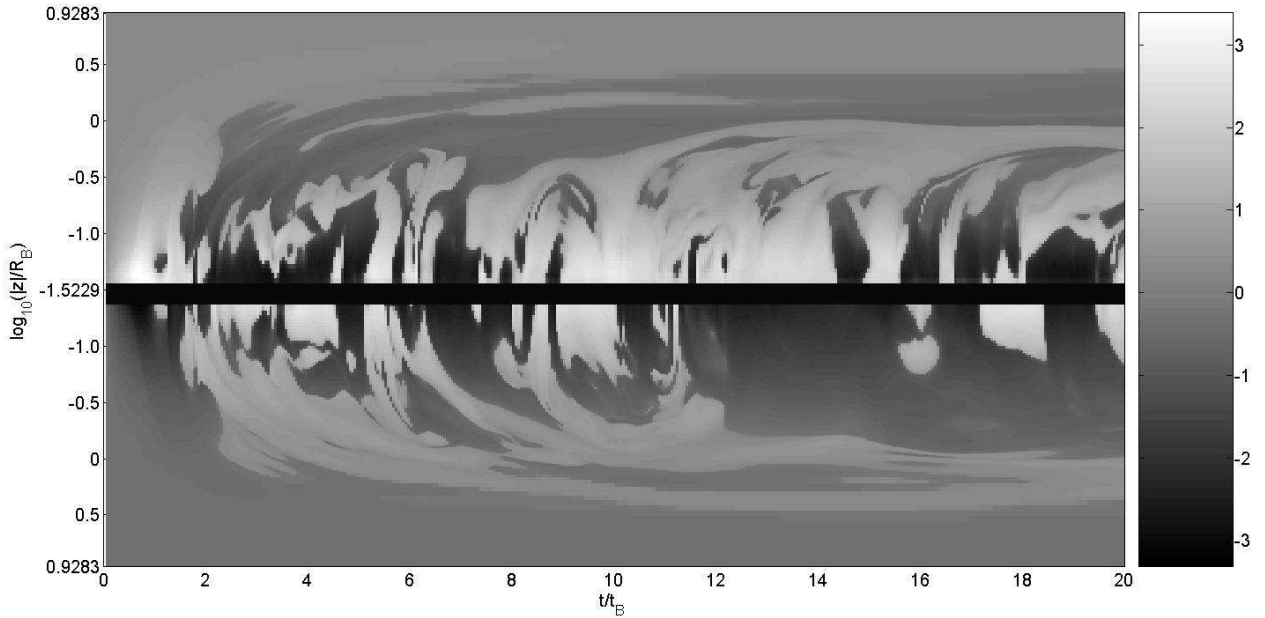}
\caption{The rotation measure integrant $\rho B_{r}$ vs
radius and time.
The central dark bar represents the inner boundary,
the vertical axis is the Z axis.
The horizontal axis is time, in units of $t_B$;
Greyscale represents $\rm{sign}(B_{r})\sqrt[4]{\rho |B_{r}|}$, which
was scaled to be more visually accessible.
The coherence time is longer at large radii and at late times.  Several
Bondi times are needed to achieve the steady state regime.
}
\label{figure:RMcellvsTime}
\end{figure*}

We can ask how well a single
measurement of RM constrains the characteristic RM, 
say the ensemble-averaged root-mean-square value RM$_{\rm rms}$.  This is a
question of how well a standard deviation is measured from a single
observation.  From figure \ref{figure:RM_PDF} we see that the distribution in our simulations 
is roughly Gaussian with standard deviation $\sigma_{\rm RM}= 0.63 {\rm RM}_{\rm est}$.
One needs to apply Bayes' Theorem to infer
the variance of a Gaussian from $N$ independent measurements:
\begin{equation}
{\Delta {\rm RM}_{\rm rms} = \left(\frac{2}{N}\right)^{1/2} \sigma_{\rm RM}}
\end{equation}
To date, no sign change in RM has been
observed, suggesting that we only have one independent measurement.
Estimating RM$_{\rm rms}$ from a single
data point requires a Bayesian inversion.  Estimating from our
simulation with a flat prior, the 95\% confidence interval for the
ensemble characteristic RM given the one data point spans 
two orders of magnitude!  

In other words, if in fact RM$_{\rm rms} = 5.4\times10^6$, it is not very surprising that we have observed 
RM$\simeq -5.4\times 10^5$.  The maximum likelihood estimate is RM$_{\rm rms} = $ RM.
The 95\% upper bound is RM$_{\rm rms} =33$RM, and the lower bound is
RM$_{\rm rms} =0.33$ RM.  More data is essential to constrain this very large
uncertainty.

A visual description of the RM integrand through the flow is shown in
figure \ref{figure:RMcellvsTime}.  The time variability time scale is
shorter at small radii, and shorter at the beginning of the
simulation.  Simulations of many Bondi times with boundaries many
Bondi radii away are necessary to see the characteristic flow patterns.

\begin{figure}
\centering
\includegraphics[scale=0.4]{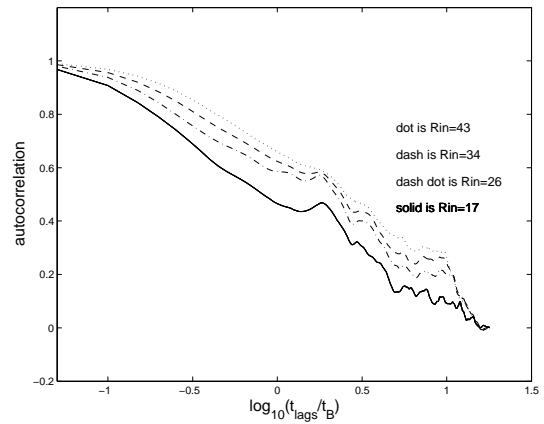}
\caption{
Autocorrelation for Figure~\protect\ref{figure:RMvsTime}.
X axis represents time lags;
Y axis represents autocorrelation for different $R_{\rm in}$.
The dotted, dashed, dashed-dot and solid lines correspond to 
$R_{\rm in}=43,34,26,17$ respectively.
}
\label{figure:auto}
\end{figure}

To be more quantitative, we plot in Figure \ref{figure:auto} the
autocorrelation of RM($t$) for different $R_{\rm rel}$.  We define the
coherence time $\tau$ to be the lag at which the autocorrelation of RM
falls to 0.5.

\begin{figure}
\centering
\includegraphics[scale=0.4]{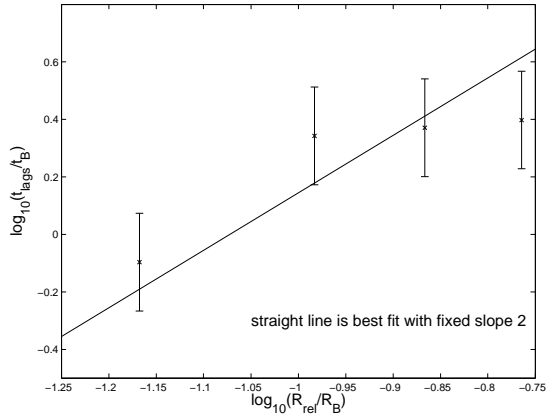}
\caption{RM coherence time $\tau$ as a function of the inner truncation radius $R_{\rm rel}$; points refer to $R_{\rm rel}=17$, $26$, $34$ and $43$.
The bootstrap error of 0.17 dex is based on the six data, two for each coordinate direction, at each $R_{\rm rel}$.
The normalization for $R_{\rm rel}={R_B}$ is 
$\log_{10}(t_{\rm lags}/t_B)=2.15$.}
\label{figure:linear_fitvstimelags}
\end{figure}

The actual RM radius $R_{\rm rel}$ is not resolved in our simulations.
In order to extrapolate to physically interesting regimes, we fit a
trend to our limited dynamic range.  The characteristic variability
time scale is given by the flow speed, so
$\tau \propto {R_{\rm rel}^3 \rho(R_{\rm rel})}/{\dot{M}}$.
For our characteristic values $k\sim 1$, we have $\tau \propto R_{\rm
  rel}^2$, which we fit to our coherence time, shown in
figure~\ref{figure:linear_fitvstimelags}.

For density profiles shallower than Bondi, the characteristic RM time
scale $\tau$ is significantly longer than the dynamical time ($\tau \sim
(R_{\rm rel}/R_B)^{3/2} t_B$).  In our fit, it is given by the accretion time 
\begin{equation}
\tau \sim 20 \left(\frac{R_{\rm rel}}{R_B}\right)^2
\left(\frac{R_{in}}{R_B}\right)^{-1/2} t_B,
\end{equation}
with a relatively large dimensionless prefactor.
This indicates a coherence time of order one year for the conditions at
Sgr A*.  The actual value of $R_{\rm rel}$ is uncertain by a factor of
6, so the expected range could be two months to a year.  

This is sufficiently distinct from one day that the
distinction between frustrated and dynamical flows should be readily
apparent, once observations span year long baselines.  We will discuss this
point more in Section \ref{obsprobes} below.

\section{Discussion} \label{S:discussion} 
In this section we wish to revisit several of the physical processes which are missing from the current numerical simulations: stellar winds from within $R_B$, the transport of energy and momentum by nearly collisionless electrons, and the inner boundary conditions imposed by a central black hole. 

{\em Stellar wind input.}  Our simulations account for the accretion of matter from outside the Bondi radius inferred from X-ray observations, but not for the direct input of matter from individual stars in the vicinity of the black hole.   \cite{2004MNRAS.350..725L} raises the possibility that individual stars may in fact dominate the accretion flow.  The wind from a single star at radius $r$ dominates the flow when its momentum output $\dot{M}_w v_w$ satisfies 
\begin{equation}\label{StarWindDominance}
\dot{M}_w v_w > 4 \pi r^2 p(r) \rightarrow 3.3 (10^{-5} M_\odot\,{\rm yr}^{-1}) (1000 \,{\rm km}\,{\rm s}^{-1}) 
\end{equation} 
where the evaluation is for a model consistent with RM constraints, in which density follows $n_H \simeq 10^{7.3} (r/R_S)^{-1} $cm$^{-3}$ and pressure follows $p \simeq 10^4 (r/R_S)^{-2}$dyn\,cm$^{-2}$ -- note that the criterion is independent of radius for $k=1$. The required momentum output, equivalent to $10^{6.2} L_\odot/c$, is well above the wind force of any of the OB stars observed within $R_B$.  While stars within $R_B$ add fresh matter faster than it is accreted by the hole, we can be confident that no single star dominates the flow. 
If the density slope is substantially more shallow, for example in
a CDAF with $k=1/2$, the steller winds would be a more important
factor.

{\em Collisionless transport.}  In the context of a dilute plasma
where Coulomb collisions are rare, electron thermal conduction has the
potential to profoundly alter the flow profile.  The importance of
this effect depends on the electrons' ability to freely stream down
their temperature gradient \citep{2008MNRAS.389.1815S}, despite the
wandering and mirroring induced by an inhomogeneous magnetic field.
The field must be weak for the magneto-thermal instability to develop,
yet weak fields are less resistive to tangling.  The thermal
conduction is expected to be strongest in the deep interior of the
flow.  If electrons actually free stream inside of 1000 Schwarzschild
radii, the electrons could be non-relativistic all the way to the
emission region, changing the interpretation of the RM.  This would
favour even shallower density profiles, for example the CDAF models.
In such a model, the RM might be expected to vary on time scales of
minutes, which appears inconsistent with current data.  If, on the
other hand, the free streaming length is short on the inside, it more
likely places the fluid in an ideal regime for the range of radii in
our simulations.  We therefore remain agnostic as to the role of
thermal conduction in hot accretion flows, although it remains a
primary caveat of the current study.  Observations of time variability
of RM will substantially improve our unstanding.

{\em Black hole inner boundary.}  Our current inner boundary
conditions do not resemble a black hole very closely, apart from the
fact that they also allow gas to accrete.  As the inner region
dominates the energetics of the flow, we consider it critical to learn
how the black hole modifies our results.  We are currently engaged in
a follow-up study with a relativistic inner boundary, to be described
in a future paper.

\subsection{Observational Probes} \label{obsprobes}

RM can be measured by several techniques.  Currently, efforts have
concentrated at high frequencies, $\nu \sim 200-300$ GHz
\citep{2006JPhCS..54..354M}, where the polarization angle varies
slowly with frequency.  Accurate measurements over long time baselines
allows discrimination between models.  At high frequencies, the SMA
and ALMA would allow a steady synaptic monitoring program.  The full
time 2 point correlation function extends the measurement space by one
dimension.

At lower frequencies, high spectral resolution is needed to
resolve the winding rate, which is now tractable with broad band high
resolution instruments such as the EVLA and ATCA/CABB.  The higher
winding rate would allow a much more sensitive measurement of small
changes in the RM, which would also be a descriminant between models.
The challenge here is that the polarization fraction drops
signficantly with frequency, requiring a more accurate instrumental
polarization model.  On the other hand, the very characteristic
$\lambda^2$ dependence of RM should allow a robust rejection against
instrumental effects.

At lower frequency, the spatial extent of the emission region is also
expected to increase.  When the emission region approaches the
rotation measure screen, one expects depolarization.  Direct polarized
VLBI imaging could shed light on this matter.  This is complicated by
interstellar scattering, which also increases the apparent angular
size.  The changing emission location as a function of frequency may
complicate the RM inferences \citep{2009ApJ...706.1353F,2010arXiv1006.5015B}
This can skew the actual value of the infered
RM, resulting in an underestimate.  The sign of RM would generically
be a more robust quantity, and looking for changes in the sign of RM
could be a proxy for the correlation function.

A separate approach is to use other polarized sources as probes of the
flow.  One candidate
population is pulsars.  At the galactic center, interstellar
scattering (Lazio 2000) smears out the pulses, making them difficult
to detect directly.  But the pulse averaged flux should still be
present.  Over the orbit around the black hole, one can measure the
time variation of the RM, leading to a probe of the spatial RM
variations in the accretion flow.  Some pulars, such as the crab,
exhibit giant pulses, which could still be visible despite a
scattering delay smearing.  These could be used to measure the
dispersion measure (DM) along the orbit.  The GMRT at 610 MHz would
have optimal sensitivity to detecting the non-pulsating emission from
pulsars, and be able to deconfuse them from the dominant synchrotron
emission using rotation measure synthesis \citep{2005A&A...441.1217B}.

\section{Conclusion}

A series of new, large dynamical range secular MHD simulation are
presented for the understanding of the low luminosity of the
supermassive black hole in the Galactic Center.  These are the first
global 3-D MHD simulations which do not face boundary conditions at
outer radii, and impose ingoing boundaries at the interior, running
for many Bondi times.  We confirm a class of magnetically frustrated
accretion flows, whose bulk properties are independent of physical and
numerical parameters, including resolution, rotation, and magnetic
fields.  No significant rotational support nor outward flow is
observed in our simulations.  An extrapolation formula is proposed and
the accretion rate is consistent with observational data.

A promising probe for the nature of the accretion flow is the rotation
measure, and its time variability.  In this comparison, the dominant
free parameter is the electron temperature.  We argued that over the
plausible range, from thermal to adiabatic, this radius varies from
40 to 250 Schwarzschild radii.  The RM variations in the
simulations are intermittent, requiring many measurements to determine
this last free parameter.

We propose that temporal rotation measure variations are a generic
prediction to distinguish between the wide variety of theoretical
models currently under consideration, ranging from CDAF through ADIOS
to ADAF.  RM is dominated by the radius at which electrons turn
relativistic, when the flow is still very subrelativistic, and is thus
much further out than the Schwarzschild radius.  Most models, other
than the ones found in our simulations, involve rapidly flowing
plasmas, with Mach numbers near unity.  These generically result in
rapid RM variations on time scales of hours to weeks (or in special
cases, it can be infinite). 
In contrast, our simulations predict variability on
time scales of weeks to years.  A major uncertainty in this
prediction is the poor statistical measure of the standard deviation
of RM measurement, which requires long term RM monitoring to quantify.

Future observations of RM time variability, or spatially resolved
measurements using pulsars, will provide valuable information.

\section{Acknowledgments}
We would like to thank Avery Broderick, 
Ramesh Narayan,
Roman Shcherbakov,
Jim Stone,
Mitch Begelman,
and Eliot Quataert for helpful discussions.  
This work was supported by the National Science and Engineering Research Council.
The work of CDM was supported by Ontario ERA.
The computations were performed on CITA's Sunnyvale clusters which are funded by the Canada Foundation for Innovation, the Ontario Innovation Trust, and the Ontario Research Fund.
The computations for the movies were performed on the GPC supercomputer at the SciNet HPC Consortium. 
SciNet is funded by: the Canada Foundation for Innovation under the auspices of Compute Canada; the Government of Ontario; Ontario Research Fund - Research Excellence; and the University of Toronto.

\bibliographystyle{mn2e}
\bibliography{bpangbib}

\begin{thebibliography}{}

\bibitem[\protect\citeauthoryear{{Agol}}{{Agol}}{2000}]{2000ApJ...538L.121A}
{Agol} E.,  2000, \apjl, 538, L121

\bibitem[\protect\citeauthoryear{{Baganoff}, {Maeda}, {Morris}, {Bautz},
  {Brandt}, {Cui}, {Doty}, {Feigelson}, {Garmire}, {Pravdo}, {Ricker} \&
  {Townsley}}{{Baganoff} et~al.}{2003}]{2003ApJ...591..891B}
{Baganoff} F.~K.,  {Maeda} Y.,  {Morris} M.,  {Bautz} M.~W.,  {Brandt} W.~N.,
  {Cui} W.,  {Doty} J.~P.,  {Feigelson} E.~D.,  {Garmire} G.~P.,  {Pravdo}
  S.~H.,  {Ricker} G.~R.,    {Townsley} L.~K.,  2003, \apj, 591, 891

\bibitem[\protect\citeauthoryear{{Blandford} \& {Begelman}}{{Blandford} \&
  {Begelman}}{1999}]{1999MNRAS.303L...1B}
{Blandford} R.~D.,  {Begelman} M.~C.,  1999, \mnras, 303, L1

\bibitem[\protect\citeauthoryear{{Bondi}}{{Bondi}}{1952}]{1952MNRAS.112..195B}
{Bondi} H.,  1952, \mnras, 112, 195

\bibitem[\protect\citeauthoryear{{Bower}, {Backer}, {Zhao}, {Goss} \&
  {Falcke}}{{Bower} et~al.}{1999}]{1999ApJ...521..582B}
{Bower} G.~C.,  {Backer} D.~C.,  {Zhao} J.-H.,  {Goss} M.,    {Falcke} H.,
  1999, \apj, 521, 582

\bibitem[\protect\citeauthoryear{{Brentjens} \& {de Bruyn}}{{Brentjens} \& {de
  Bruyn}}{2005}]{2005A&A...441.1217B}
{Brentjens} M.~A.,  {de Bruyn} A.~G.,  2005, \aap, 441, 1217

\bibitem[\protect\citeauthoryear{{Broderick} \& {Blandford}}{{Broderick} \&
  {Blandford}}{2010}]{2010ApJ...718.1085B}
{Broderick} A.~E.,  {Blandford} R.~D.,  2010, \apj, 718, 1085

\bibitem[\protect\citeauthoryear{{Broderick} \& {McKinney}}{{Broderick} \&
  {McKinney}}{2010}]{2010arXiv1006.5015B}
{Broderick} A.~E.,  {McKinney} J.~C.,  2010, ArXiv e-prints

\bibitem[\protect\citeauthoryear{{Fish}, {Doeleman}, {Broderick}, {Loeb} \&
  {Rogers}}{{Fish} et~al.}{2009}]{2009ApJ...706.1353F}
{Fish} V.~L.,  {Doeleman} S.~S.,  {Broderick} A.~E.,  {Loeb} A.,    {Rogers}
  A.~E.~E.,  2009, \apj, 706, 1353

\bibitem[\protect\citeauthoryear{{Genzel}, {Sch{\"o}del}, {Ott}, {Eisenhauer},
  {Hofmann}, {Lehnert}, {Eckart}, {Alexander}, {Sternberg}, {Lenzen},
  {Cl{\'e}net}, {Lacombe}, {Rouan}, {Renzini} \& {Tacconi-Garman}}{{Genzel}
  et~al.}{2003}]{2003ApJ...594..812G}
{Genzel} R.,  {Sch{\"o}del} R.,  {Ott} T.,  {Eisenhauer} F.,  {Hofmann} R.,
  {Lehnert} M.,  {Eckart} A.,  {Alexander} T.,  {Sternberg} A.,  {Lenzen} R.,
  {Cl{\'e}net} Y.,  {Lacombe} F.,  {Rouan} D.,  {Renzini} A.,
  {Tacconi-Garman} L.~E.,  2003, \apj, 594, 812

\bibitem[\protect\citeauthoryear{{Gillessen}, {Eisenhauer}, {Trippe},
  {Alexander}, {Genzel}, {Martins} \& {Ott}}{{Gillessen}
  et~al.}{2009}]{2009ApJ...692.1075G}
{Gillessen} S.,  {Eisenhauer} F.,  {Trippe} S.,  {Alexander} T.,  {Genzel} R.,
  {Martins} F.,    {Ott} T.,  2009, \apj, 692, 1075

\bibitem[\protect\citeauthoryear{{Gruzinov}}{{Gruzinov}}{2001}]{2001astro.ph..%
4113G}
{Gruzinov} A.,  2001, ArXiv Astrophysics e-prints

\bibitem[\protect\citeauthoryear{{Igumenshchev} \& {Narayan}}{{Igumenshchev} \&
  {Narayan}}{2002}]{2002ApJ...566..137I}
{Igumenshchev} I.~V.,  {Narayan} R.,  2002, \apj, 566, 137

\bibitem[\protect\citeauthoryear{{Igumenshchev}, {Narayan} \&
  {Abramowicz}}{{Igumenshchev} et~al.}{2003}]{2003ApJ...592.1042I}
{Igumenshchev} I.~V.,  {Narayan} R.,    {Abramowicz} M.~A.,  2003, \apj, 592,
  1042

\bibitem[\protect\citeauthoryear{{Johnson} \& {Quataert}}{{Johnson} \&
  {Quataert}}{2007}]{2007ApJ...660.1273J}
{Johnson} B.~M.,  {Quataert} E.,  2007, \apj, 660, 1273

\bibitem[\protect\citeauthoryear{{Kaeppeli}, {Whitehouse}, {Scheidegger}, {Pen}
  \& {Liebendoerfer}}{{Kaeppeli} et~al.}{2009}]{2009arXiv0910.2854K}
{Kaeppeli} R.,  {Whitehouse} S.~C.,  {Scheidegger} S.,  {Pen} U.,
  {Liebendoerfer} M.,  2009, ArXiv e-prints

\bibitem[\protect\citeauthoryear{{Levin} \& {Beloborodov}}{{Levin} \&
  {Beloborodov}}{2003}]{2003ApJ...590L..33L}
{Levin} Y.,  {Beloborodov} A.~M.,  2003, \apjl, 590, L33

\bibitem[\protect\citeauthoryear{{Loeb}}{{Loeb}}{2004}]{2004MNRAS.350..725L}
{Loeb} A.,  2004, \mnras, 350, 725

\bibitem[\protect\citeauthoryear{{Marrone}, {Moran}, {Zhao} \& {Rao}}{{Marrone}
  et~al.}{2006}]{2006JPhCS..54..354M}
{Marrone} D.~P.,  {Moran} J.~M.,  {Zhao} J.-H.,    {Rao} R.,  2006, Journal of
  Physics Conference Series, 54, 354

\bibitem[\protect\citeauthoryear{{Marrone}, {Moran}, {Zhao} \& {Rao}}{{Marrone}
  et~al.}{2007}]{2007ApJ...654L..57M}
{Marrone} D.~P.,  {Moran} J.~M.,  {Zhao} J.-H.,    {Rao} R.,  2007, \apjl, 654,
  L57

\bibitem[\protect\citeauthoryear{{Narayan}, {Igumenshchev} \&
  {Abramowicz}}{{Narayan} et~al.}{2000}]{2000ApJ...539..798N}
{Narayan} R.,  {Igumenshchev} I.~V.,    {Abramowicz} M.~A.,  2000, \apj, 539,
  798

\bibitem[\protect\citeauthoryear{{Narayan} \& {Yi}}{{Narayan} \&
  {Yi}}{1994}]{1994ApJ...428L..13N}
{Narayan} R.,  {Yi} I.,  1994, \apjl, 428, L13

\bibitem[\protect\citeauthoryear{{Pang}, {Pen} \& {Perrone}}{{Pang}
  et~al.}{2010}]{2010arXiv1004.1680P}
{Pang} B.,  {Pen} U.,    {Perrone} M.,  2010, ArXiv e-prints

\bibitem[\protect\citeauthoryear{{Pen}, {Arras} \& {Wong}}{{Pen}
  et~al.}{2003}]{2003ApJS..149..447P}
{Pen} U.-L.,  {Arras} P.,    {Wong} S.,  2003, \apjs, 149, 447

\bibitem[\protect\citeauthoryear{{Pen}, {Matzner} \& {Wong}}{{Pen}
  et~al.}{2003}]{2003ApJ...596L.207P}
{Pen} U.-L.,  {Matzner} C.~D.,    {Wong} S.,  2003, \apjl, 596, L207

\bibitem[\protect\citeauthoryear{{Proga} \& {Begelman}}{{Proga} \&
  {Begelman}}{2003}]{2003ApJ...592..767P}
{Proga} D.,  {Begelman} M.~C.,  2003, \apj, 592, 767

\bibitem[\protect\citeauthoryear{{Quataert} \& {Gruzinov}}{{Quataert} \&
  {Gruzinov}}{2000a}]{2000ApJ...545..842Q}
{Quataert} E.,  {Gruzinov} A.,  2000a, \apj, 545, 842

\bibitem[\protect\citeauthoryear{{Quataert} \& {Gruzinov}}{{Quataert} \&
  {Gruzinov}}{2000b}]{2000ApJ...539..809Q}
{Quataert} E.,  {Gruzinov} A.,  2000b, \apj, 539, 809

\bibitem[\protect\citeauthoryear{{Quataert} \& {Narayan}}{{Quataert} \&
  {Narayan}}{1999}]{1999ApJ...520..298Q}
{Quataert} E.,  {Narayan} R.,  1999, \apj, 520, 298

\bibitem[\protect\citeauthoryear{{Revnivtsev}, {Churazov}, {Sazonov},
  {Sunyaev}, {Lutovinov}, {Gilfanov}, {Vikhlinin}, {Shtykovsky} \&
  {Pavlinsky}}{{Revnivtsev} et~al.}{2004}]{2004A&A...425L..49R}
{Revnivtsev} M.~G.,  {Churazov} E.~M.,  {Sazonov} S.~Y.,  {Sunyaev} R.~A.,
  {Lutovinov} A.~A.,  {Gilfanov} M.~R.,  {Vikhlinin} A.~A.,  {Shtykovsky}
  P.~E.,    {Pavlinsky} M.~N.,  2004, \aap, 425, L49

\bibitem[\protect\citeauthoryear{{Sharma}, {Quataert} \& {Stone}}{{Sharma}
  et~al.}{2007}]{2007ApJ...671.1696S}
{Sharma} P.,  {Quataert} E.,    {Stone} J.~M.,  2007, \apj, 671, 1696

\bibitem[\protect\citeauthoryear{{Sharma}, {Quataert} \& {Stone}}{{Sharma}
  et~al.}{2008}]{2008MNRAS.389.1815S}
{Sharma} P.,  {Quataert} E.,    {Stone} J.~M.,  2008, \mnras, 389, 1815

\bibitem[\protect\citeauthoryear{{Shcherbakov}}{{Shcherbakov}}{2008}]{2008ApJ.%
..688..695S}
{Shcherbakov} R.~V.,  2008, \apj, 688, 695

\bibitem[\protect\citeauthoryear{{Shcherbakov} \& {Baganoff}}{{Shcherbakov} \&
  {Baganoff}}{2010}]{2010arXiv1004.0702S}
{Shcherbakov} R.~V.,  {Baganoff} F.~K.,  2010, ArXiv e-prints

\bibitem[\protect\citeauthoryear{{Tanaka} \& {Menou}}{{Tanaka} \&
  {Menou}}{2006}]{2006ApJ...649..345T}
{Tanaka} T.,  {Menou} K.,  2006, \apj, 649, 345

\bibitem[\protect\citeauthoryear{{Yuan}, {Quataert} \& {Narayan}}{{Yuan}
  et~al.}{2003}]{2003ApJ...598..301Y}
{Yuan} F.,  {Quataert} E.,    {Narayan} R.,  2003, \apj, 598, 301

\bibitem[\protect\citeauthoryear{{Yusef-Zadeh}, {Bushouse}, {Wardle} \& {11
  coauthors}}{{Yusef-Zadeh} et~al.}{2009}]{2009arXiv0907.3786Y}
{Yusef-Zadeh} F.,  {Bushouse} H.,  {Wardle} M.,    {11 coauthors} 2009, ArXiv
  e-prints

\bibitem[\protect\citeauthoryear{{Yusef-Zadeh}, {Bushouse}, {Wardle}, {Heinke},
  {Roberts}, {Dowell}, {Brunthaler}, {Reid}, {Martin}, {Marrone}, {Porquet},
  {Grosso}, {Dodds-Eden}, {Bower}, {Wiesemeyer} \& {Miyazaki}}{{Yusef-Zadeh}
  et~al.}{2009}]{2009ApJ...706..348Y}
{Yusef-Zadeh} F.,  {Bushouse} H.,  {Wardle} M.,  {Heinke} C.,  {Roberts} D.~A.,
   {Dowell} C.~D.,  {Brunthaler} A.,  {Reid} M.~J.,  {Martin} C.~L.,  {Marrone}
  D.~P.,  {Porquet} D.,  {Grosso} N.,  {Dodds-Eden} K.,  {Bower} G.~C.,
  {Wiesemeyer} H.,    {Miyazaki} A.,  2009, \apj, 706, 348

\bibitem[\protect\citeauthoryear{{Yusef-Zadeh}, {Muno}, {Wardle} \&
  {Lis}}{{Yusef-Zadeh} et~al.}{2007}]{2007ApJ...656..847Y}
{Yusef-Zadeh} F.,  {Muno} M.,  {Wardle} M.,    {Lis} D.~C.,  2007, \apj, 656,
  847

\end{thebibliography}

\appendix

\section{Rotation measure constraint on accretion flow}\label{A:RMconstraint} 

In traversing the accretion flow, linearly polarized radio waves of wavelength $\lambda$ are rotated by RM\,$\lambda^2$ radians, where 
\begin{equation} \label{RM}
\mbox{RM} = {e^3  \over2\pi m_e^2 c^4} \int n_e f(T_e) B \cos(\theta)    dl. 
\end{equation} 
Here $f(T_e)$ is a ratio of modified Bessel functions: $f(T_e) = K_0(m_e c^2/k_B T_e)/K_2(m_e c^2/k_B T_e)$ \citep{2008ApJ...688..695S}, which suppresses RM by a factor $\propto T_e^{-2}$ wherever electrons are relativistic.  The integral here covers the entire path from source to observer; $\theta$ is the angle between $\mathbf B$ and the line of sight.    This expression is appropriate for the frequencies at which RM has been observed; at lower frequencies, where propagation is ``superadiabatic'' \citep{2010ApJ...718.1085B} $\cos(\theta)\rightarrow\pm |\cos(\theta)|$. 

We adopt a power-law solution with negligible rotational support in which $\rho\propto r^{-k}$, and the total pressure $P\propto r^{-k_P}$ with $k_P=k+1$; moreover we take $T_e\propto r^{-k_T}$ for the relativistic electrons.    The hydrostatic equation $dP/dr = -G M\rho/r^2$ becomes 
\begin{equation} 
P = P_g + P_B = {G M \over  (k+1)} {\rho\over r}, 
\end{equation}
and with $P_g = \beta P_B = \beta B^2/(8\pi)$, $\rho = n_e \mu_e$ (where $\mu_e = 1.2 m_p$ is the mass per electron),
\begin{equation} 
B = \left[{8\pi \over (\beta+1)(k+1) } {G M \mu_e n_e \over r} \right]^{1/2}.
\end{equation} 

\begin{figure}
\centering
\includegraphics[width=8.4cm]{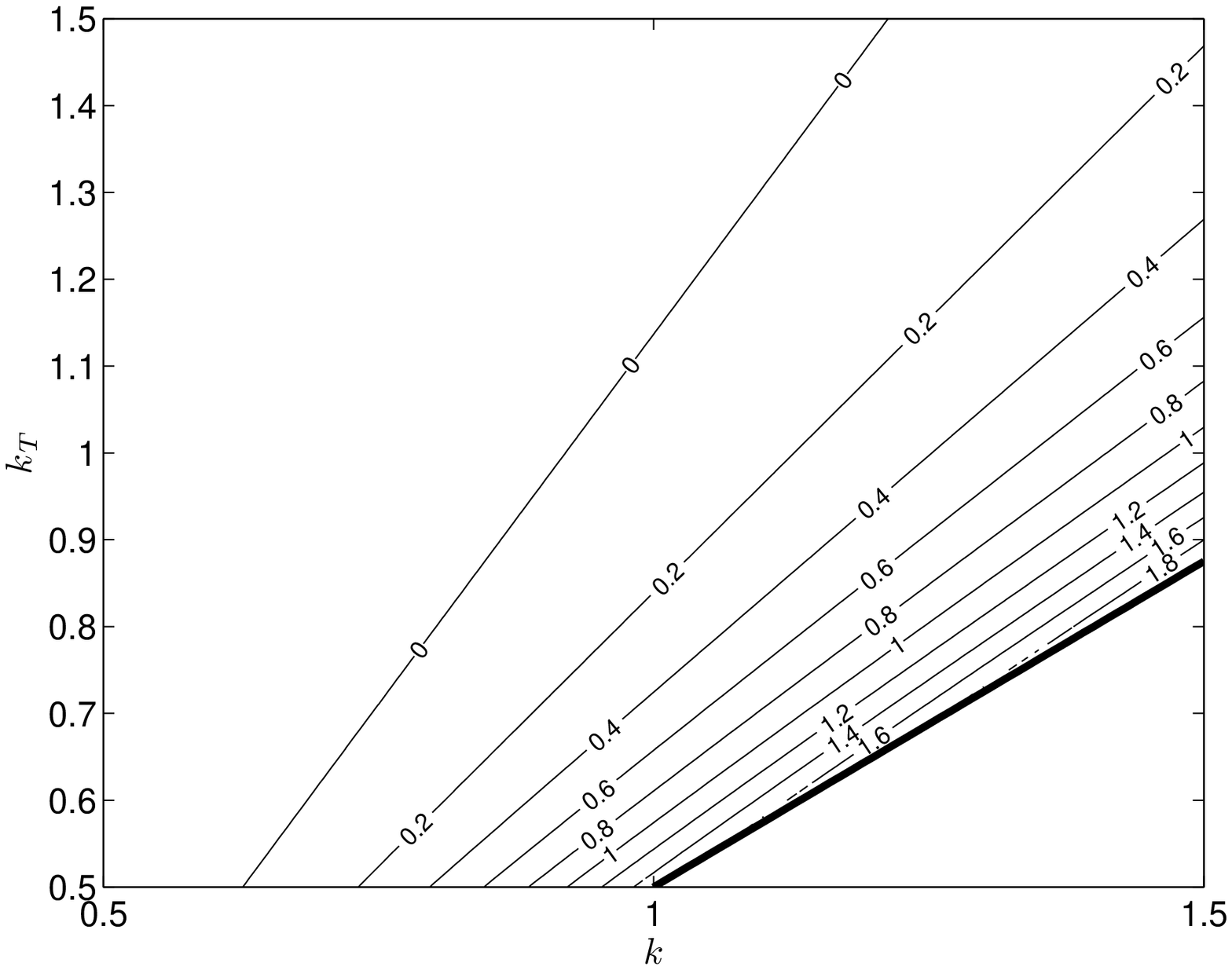}
\caption{The logarithm of the relativistic RM factor,  $\log_{10} F(k,k_T)$.  The true RM integral is modified by a factor $F(k,k_T)$ relative to an estimate in which the nonrelativistic formula is used, but the inner bound of integration is set to the radius $R_{\rm rel}$ at which electrons become relativistic; see equation \ref{RM}. } 
\label{fig:Ffactort}
\end{figure}

So long as $k>1/3$ (so that RM converges at large radii) and $k<(1+4k_T)/3$ (so it converges inward as well), the RM integral is set around  $R_{\rm rel}$.  
Taking a radial line of sight ($dl \rightarrow dr$), we write 
\begin{eqnarray} 
 \int_0^\infty n_e f(T_e) B \cos(\theta) dr &=&  F(k, k_T) \int_{R_{\rm rel}}^\infty n_e B \cos(\theta) dr \\  &=&   {2\over3k-1} \left<\cos(\theta)\right> F(k,k_T)  \left[ n_e B r\right]_{R_{\rm rel}} \nonumber 
\end{eqnarray} 
where  $\left<\cos(\theta)\right>$ encapsulates the difference between the true integral what it would have been if $\theta=0$ all along the path, and $F(k,k_T)$  encapsulates the difference between a smooth cutoff and a sharp one.   We plot $F(k,k_T)$ in Figure \ref{fig:Ffactort}; it is of order unity except as $k_T$ approaches $(3k-1)/4$.   All together, 
\begin{equation} \label{RMform} 
\mbox{RM} = {4e^2 G M\over m_e^2 c^5} { \left<\cos(\theta)\right> F(k,k_T) \over 3k-1} \left[{ \mu_e n_e(R_{\rm rel})^3 
\over \pi (k+1)(\beta+1) }  {R_{\rm rel}\over R_S}\right]^{1/2} . 
\end{equation}

To estimate $n_e(R_{\rm rel})$ from RM, one must make assumptions about the uncertain parameters $\beta$, $\left<\cos(\theta)\right>$, $k_T$, and $R_{\rm rel}/R_S$; then $k$ can be derived self-consistently from observations $n_e(R_B)$ and RM.   Our fiducial values of these parameters are 10, 0.5, 0.5 and 100, respectively, of which we consider the last to be the most uncertain.  We now discuss each in turn. 

Although the magnetization parameter $\beta$ could conceivably take a very wide range of values, we consistently find $\beta\simeq 10$ in our simulations, with some tendency for $\beta$ to decrease inward.   We consider it unlikely for the flow to be much less magnetized, given the magnetization of the galactic center and the fact that weak fields are enhanced in most of the flow models under consideration. 

If $\mathbf B$ wanders little in the region where the integrand is large (a zone of width $\sim R_{\rm rel}$ around $R_{\rm rel}$), and is randomly oriented relative to the line of sight $\left<\cos(\theta)\right>\simeq \cos(\theta(R_{\rm rel}))$, typically $1/2$ in absolute value.   If the field were purely radial, $\left<\cos(\theta)\right>$ would be unity.  Conversely if $\mathbf B$ reverses frequently in this region (the number of reversals $N_r$ is large) then $\left<\cos(\theta)\right>_{\rm rms} \simeq 1/(2\sqrt{N_r+1})$ will be small.    However, $N_r$ cannot be too large, or magnetic forces are unbalanced.  We gauge its maximum value by equating the square of the buoyant growth rate, ${\cal N}^2 = [(3-2k)/5] GM/r^3$, against the square of the Alfv\'en frequency $N_r^2 v_A^2/r^2$. Noting that $v_A^2 = G M/[2(\beta+1)(k+1) r]$, we find $N_r^2 \simeq (2/5) (\beta+1)(k+1)(3-2k)$.  For $\beta=10$ and $k=1$ this implies $\left<\cos(\theta)\right>_{\rm rms} \simeq 0.25$: a very minor suppression.   We can therefore be confident that $\left<\cos(\theta)\right> = 0.5$ to within a factor of 2, unless $\beta\gg 10$ for some reason. 

The precise value of $k_T$ is not important unless it approaches or falls below the minimum value $(3k-1)/4$.  If electron conduction is very strong this is unavoidable, as rapid transport implies $k_T \simeq 0$; however in this case the relativistic region disappears, as discussed below.  Alternately, if relativistic electrons are trapped and adiabatic, $T_e\propto \rho^{1/3}$ and $k_T=k/3$; however $k_T<(3k-1)/4$ then requires $k<3/5$, which can only be realized within the CDAF model.  Finally, if electrons remain strongly coupled to ions, $k_T =1$ and we only require $k<5/3$. 

The location at which electrons become relativistic, $R_{\rm rel}$, is quite uncertain.  Models such as those of \citet{2003ApJ...598..301Y}, in which electrons are heated while advecting inward, predict $R_{\rm rel}\simeq 10^2 R_S$.   The maximum conceivable $R_{\rm rel}$ corresponds to adiabatic compression of the electrons, inward from the radius at which they decouple from ions; this yields about $500/(1+k) R_S$.   If conduction is very strong, however, electrons should remain cold throughout the flow; in this case we should replace $R_{\rm rel}/R_S\rightarrow 1$ and $F(k,k_T)\rightarrow 1$ in equation (\ref{RMform}).   

Adopting our fiducial values for the other variables, and taking $F(k,k_T)\rightarrow 1$ for lack of knowledge regarding $k_T$, we may solve for the self-consistent value of $k$ which connects the density at $R_B$ with $n_e(R_{\rm rel})$ derived from equation \ref{RMform}.   We find $k\rightarrow (0.90, 1.23, 1.32)$ for $R_{\rm rel}/R_S \rightarrow (200,100, 1)$, respectively.   
As noted in the text, the current small set of RM measurements allows
a two order of magnitude range in RM$_{\rm est}$, and $k\sim 1$ is consistent
with data.  Longer observations of time and amplitude will improve
the constraints.

\section{Inner boundary conditions}\label{A:InnerBCs} 

The inner boundary conditions were determined by first solving for the
vacuum solution of the magnetic field inside the entire inner boundary
cube.  Then inside the largest possible sphere within this cube,
matter and energy were removed.

To simplify the programming, we put the entire inner boundary region
on one node.  This meant that the grid had to be divided over an odd
number of nodes in each Cartesian direction.

\subsection{Magnetic field} \label{A:Inner-B} 

In order to determine the vacuum magnetic field solution, we use the
following two Maxwell equations for zero current:
\begin{equation} \label{divb}
\nabla \cdot \bmath{B} = 0 \textrm{ ,}
\end{equation}
\begin{equation} \label{curlb}
\nabla \times \bmath{B} = 0 \textrm{ .}
\end{equation}
Equation (\ref{curlb}) enables us to write $\bmath{B}=\nabla \phi$,
for some scalar function $\phi$.  Combining this with (\ref{divb}) we
obtain Laplace's equation
\begin{equation} \label{laplace}
\nabla^2\phi=0 \textrm{ ,}
\end{equation}
which we solve with Neumann boundary conditions (the normal derivative
$\hat{\bmath{n}} \cdot \nabla \phi $ specified) given by $\bmath{B}
\cdot \hat{\bmath{n}}$ on the boundary of the cube.

Since the MHD code stores the values of $\bmath{B}$ on the left-hand
cell faces, we must solve for $\phi$ in cell centers and then take
derivatives to get the value of $\bmath{B}$ on the cell boundary.  Let
the inner boundary cube be of side length $N$, consisting of cells
numbered $1,\ldots,N$ in all three directions.  In order to simplify
the problem we set $\bmath{B} \cdot \hat{\bmath{n}}=0$ on five of the
six faces of the cube, and find the contribution to $\phi$ from one
face at a time.

Suppose $\bmath{B} \cdot \hat{\bmath{n}}=0$ on all of the faces except
the $i=N+1$ face (i.e., $B_x^{N+1,j,k}$ can be non-zero).  The Laplace
equation (\ref{laplace}) with Neumann boundary conditions only has a
solution if the net flux of field into the cube is zero.  Since all of the
boundary conditions are zero except for the $i=N+1$ face, that face
must have a net flux through it of zero.  Defining
\begin{equation}
\overline{B^{N+1}_x}=\frac{1}{N^2}
\sum_{j=1}^N \sum_{k=1}^N B^{N+1,j,k}_x
\end{equation}
to be the average of $B_x$ on the $i=N+1$ face, and letting
$b^{N+1,j,k}_x=B^{N+1,j,k}_x- \overline{B^{N+1}_x}$, $b_x^{N+1,j,k}$
can be used as the boundary condition and $\overline{B^{N+1}_x}$ will
be added in later.

We use separation of variables to solve for $\phi$.  Set
$\phi^{ijk}=X^i Y^j Z^k$, substitute into (\ref{laplace}), and
rearrange to get {\setlength\arraycolsep{2pt}
\begin{eqnarray} \label{laplace2}
\frac{X^{i+1}-2 X^i+X^{i-1}}{X^i}+\frac{Y^{j+1}-2 Y^j+Y^{j-1}}{Y^j}+&
&\nonumber\\ +\frac{Z^{k+1}-2 Z^k+Z^{k-1}}{Z^k}&=&0 \textrm{ .}
\end{eqnarray}}
Now let
\begin{equation} \label{yeqn}
\frac{Y^{j+1}-2 Y^j+Y^{j-1}}{Y^j}=-\eta^2 \textrm{ , and}
\end{equation}
\begin{equation} \label{zeqn}
\frac{Z^{k+1}-2 Z^k+Z^{k-1}}{Z^k}=-\omega^2 \textrm{ .}
\end{equation}
Solving equations (\ref{yeqn}) and (\ref{zeqn}) with the boundary
conditions,
\begin{equation}
Y^j_m=\cos{\frac{m\pi(j-\frac{1}{2})}{N}} \textrm{ ,}\qquad
Z^k_n=\cos{\frac{n\pi(k-\frac{1}{2})}{N}} \textrm{ ,}
\end{equation}
\begin{equation} \label{etaomega}
\eta_m^2=4\sin^2{\frac{m\pi}{2N}} \textrm{ ,}\qquad 
\omega_n^2=4\sin^2{\frac{n\pi}{2N}} \textrm{ .}
\end{equation}
Substituting (\ref{yeqn}), (\ref{zeqn}), and (\ref{etaomega}) into
(\ref{laplace2}), and solving, yields
\begin{equation}
X^i_{mn}=\cosh{\frac{\alpha_{mn} \pi (i-\frac{1}{2})}{2N}} \textrm{ ,}
\end{equation}
where
\begin{equation}
\alpha_{mn} = \frac{2N}{\pi} \textrm{arcsinh}{\sqrt{\sin^2{\frac{n\pi}{2N}}+
\sin^2{\frac{m\pi}{2N}}}} \textrm{ .}
\end{equation}
Finally, putting this all together,
{\setlength\arraycolsep{2pt}
\begin{eqnarray}
\phi^{ijk}&=&\sum_{m=0}^{N-1}\sum_{n=0}^{N-1} A_{mn}
\cos{\frac{m\pi(j-\frac{1}{2})}{N}}
\cos{\frac{n\pi(k-\frac{1}{2})}{N}} \times \nonumber \\ & & \times
\cosh{\frac{\alpha_{mn}\pi(i-\frac{1}{2})}{N}} \textrm{ ,}
\end{eqnarray}}
and define $A_{00}=0$.

To determine the coefficients $A_{mn}$ we add in the final boundary
condition ($i=N+1$), and get
{\setlength\arraycolsep{2pt}
\begin{eqnarray}
A_{mn}&=&\frac{4}{N^2} \frac{1}{2\sinh(\alpha_{mn}\pi)\sinh(\alpha_{mn}\pi/2N)}
\times \nonumber \\ & & \times
\sum_{j=1}^N \sum_{k=1}^N b_x^{N+1,j,k}
\cos{\frac{m\pi(j-\frac{1}{2})}{N}} \times \nonumber \\ & & \times
\cos{\frac{n\pi(k-\frac{1}{2})}{N}} \textrm{ .}
\end{eqnarray}}
A similar calculation may be performed for the case when the $i=1$
boundary has non-zero field.  After finding the contribution from each
face, store their sum in $\phi$.

To deal with the subtracted cube face field averages, let
{\setlength\arraycolsep{2pt}
\begin{eqnarray}
\phi_0^{ijk} &=& \overline{B_x^{1jk}}i + \overline{B_y^{i1k}}j +
\overline{B_z^{ij1}}k + \nonumber \\ & & +
\frac{\overline{B_x^{N+1,j,k}}-\overline{B_x^{1jk}}}{2N}(i^2+j^2)+
\nonumber \\ & & +
\frac{\overline{B_x^{N+1,j,k}}+\overline{B_y^{i,N+1,k}}-\overline{B_x^{1jk}}
-\overline{B_y^{i1k}}}{2N}(j^2+k^2) \textrm{,}
\end{eqnarray}}
and add this to $\phi$.  $\phi_0$ is the potential of a cube where
each face has the uniform magnetic field given by the average of the
magnetic field on the corresponding face of the inner boundary cube.

To find $\bmath{B}$, set
{\setlength\arraycolsep{2pt}
\begin{eqnarray}
B_x^{ijk}&=&\phi^{ijk}-\phi^{i-1,j,k} \textrm{ ,} \\
B_y^{ijk}&=&\phi^{ijk}-\phi^{i,j-1,k} \textrm{ ,} \\
B_z^{ijk}&=&\phi^{ijk}-\phi^{i,j,k-1} \textrm{ .}
\end{eqnarray}}

\begin{figure}
\centering
\includegraphics[width=8.4cm]{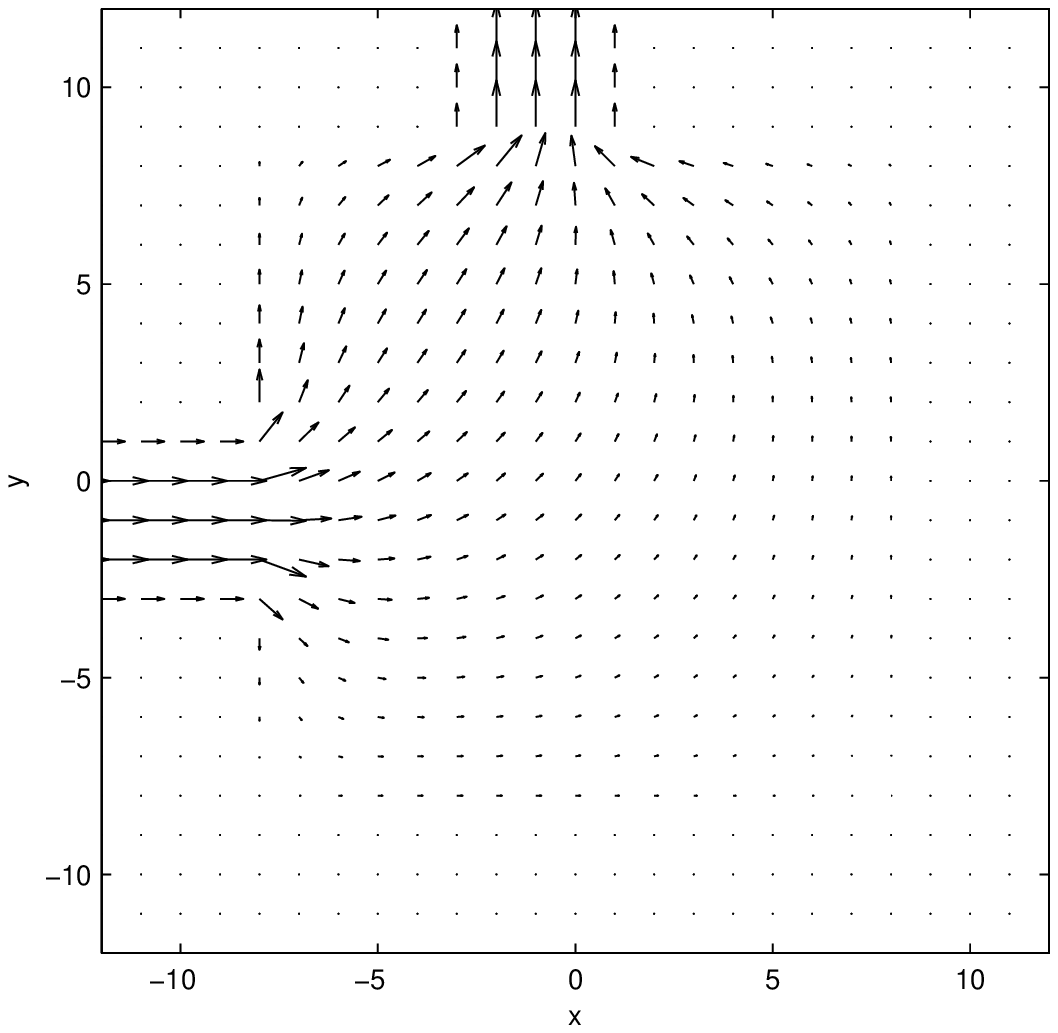}
\caption{Vacuum solution of the magnetic field is calculated in the
central region.  The field lines outside of the central region show
the boundary condition.} \label{fig:btest}
\end{figure}

In Figure \ref{fig:btest} we used the magnetic field solver with a
boundary condition consisting of field going in one side and out an
adjacent side of the box.  This boundary condition tests both the
$\phi_0$ component of the solution (since faces have non-zero net
flux) as well as the Fourier series component (since faces have
non-constant magnetic field).

\subsection{Density and pressure}\label{A:Inner-gas} 

Inside the largest possible sphere that can be inscribed within the
inner boundary cube, we adjust the density and pressure so that the
Alfv\'en speed and the sound speed are both equal to the circular
speed.  We accomplish this by setting
\begin{equation}
\rho=\frac{\bmath{B}^2 r}{G M_{BH}} \textrm{ ,}
\end{equation}
\begin{equation}
p=\frac{G M_{BH} \rho}{r \gamma} \textrm{ .}
\end{equation}
We then set $p$ to $0.1p$.  $\rho$ and $p$ were assigned minimum
values of $0.1$ times the average value of $\rho$ outside of the
sphere, and $0.001$, respectively, to ensure stability.

\bsp

\section{Supporting Information}

{\bf Movie 1}. Animation of magnetically frustrated convection simulation.

The qualitative behavious of the accretion flow is best illustrated
in the form of a movie.  This movie shows case 25.
The raw simulation used $600^3$ grid cells.
The Bondi radius is at $1000$ grid units, where one grid unit
is the smallest central grid spacing.  The full box size is $8000^3$
grid units. Colour represents the entropy,
and arrows represent the magnetic field vector.
The right side shows the equatorial plane (yz).
the left side shows a perpendicular plane (xy).
The moving white circles represent the flow of an unmagnetized
Bondi solution, starting at the Bondi radius.
On average, the fluid is slowly moving inward, in a state of magnetically 
frustrated convection.
Various other 
formats can also be seen at 
http://www.cita.utoronto.ca/~pen/MFAF/blackhole{\_}movie/index.html.

\label{lastpage}

\end{document}